\input{psfig}
\documentstyle[12pt]{article}
\title{ \normalsize  \bf  PERFORMANCE AND STABILITY OF A WINGED VEHICLE IN GROUND EFFECT}
\author{{ { Nicola de Divitiis}\thanks{Department of Mechanics and Aeronautics, via Eudossiana, 18, 00184
} }
 \\ {\it  University of Rome ``La Sapienza'',}
{ \it  Rome, Italy
}
} \date{}

\begin{document}
\baselineskip=0.950cm
\thispagestyle{empty}

\maketitle


\newcommand{\no}{\noindent}
\newcommand{\be}{\begin{equation}}
\newcommand{\ee}{\end{equation}}
\newcommand{\bea}{\begin{eqnarray}}
\newcommand{\eea}{\end{eqnarray}}
\newcommand{\bc}{\begin{center}}
\newcommand{\ec}{\end{center}}
\newcommand{\ds}{\displaystyle}

\newcommand{\calr}{{\cal R}}
\newcommand{\calv}{{\cal V}}

\newcommand{\bff}{\mbox{\boldmath $f$}}
\newcommand{\bfg}{\mbox{\boldmath $g$}}
\newcommand{\bfh}{\mbox{\boldmath $h$}}
\newcommand{\bfi}{\mbox{\boldmath $i$}}
\newcommand{\bfm}{\mbox{\boldmath $m$}}
\newcommand{\bfp}{\mbox{\boldmath $p$}}
\newcommand{\bfr}{\mbox{\boldmath $r$}}
\newcommand{\bfu}{\mbox{\boldmath $u$}}
\newcommand{\bfv}{\mbox{\boldmath $v$}}
\newcommand{\bfx}{\mbox{\boldmath $x$}}
\newcommand{\bfy}{\mbox{\boldmath $y$}}
\newcommand{\bfw}{\mbox{\boldmath $w$}}
\newcommand{\bfk}{\mbox{\boldmath $k$}}

\newcommand{\bfA}{\mbox{\boldmath $A$}}
\newcommand{\bfD}{\mbox{\boldmath $D$}}
\newcommand{\bfI}{\mbox{\boldmath $I$}}
\newcommand{\bfL}{\mbox{\boldmath $L$}}
\newcommand{\bfM}{\mbox{\boldmath $M$}}
\newcommand{\bfS}{\mbox{\boldmath $S$}}
\newcommand{\bfT}{\mbox{\boldmath $T$}}
\newcommand{\bfU}{\mbox{\boldmath $U$}}
\newcommand{\bfX}{\mbox{\boldmath $X$}}
\newcommand{\bfY}{\mbox{\boldmath $Y$}}
\newcommand{\bfK}{\mbox{\boldmath $K$}}

\newcommand{\bflambda}{\mbox{\boldmath $\lambda$}}
\newcommand{\bfrho}{\mbox{\boldmath $\rho$}}
\newcommand{\bfomega}{\mbox{\boldmath $\omega$}}
\newcommand{\bfeps}{\mbox{\boldmath $\varepsilon$}}
\newcommand{\bfphi}{\mbox{\boldmath $\Phi$}}
\newcommand{\bfOmega}{\mbox{\boldmath $\Omega$}}
\newcommand{\bint}{\mbox{ \int{a}{b}} }



\bc
 {\bf Abstract}
\ec

Present work deals with the dynamics of vehicles which intentionally operate in
the ground proximity.
The dynamics in ground effect is influenced by the vehicle orientation with respect to the ground,
since  the aerodynamic force and moment coefficients, which in turn depend on height and angle of attack,
also vary with the Euler angles. This feature, usually neglected in the applications, can be
responsible for sizable variations of the aircraft performance and stability.
A further effect, caused by the sink rate, determines unsteadiness that modifies the aerodynamic coefficients. In this work, an analytical formulation is proposed
for the force and moment calculation in the presence of the ground and taking the
aircraft attitude and sink rate into account.
The aerodynamic coefficients are firstly calculated for a representative vehicle
and its characteristics in ground effect are investigated.
Performance and  stability characteristics are then discussed with reference to significant
equilibrium conditions, while the non-linear dynamics is studied through the numerical
integration of the equations of motion.

\bc
{\bf Nomenclature}
\ec
{
\no \begin{tabular}{ll}
\end{tabular}
}

{
\no \begin{tabular}{ll}
 $AR $                & aspect ratio \\\\
 $b$                  & wing span \\\\
 $c$                  & mean wing chord  \\\\
 $C_D$                & drag coefficient \\\\
  $C_L$                & lift coefficient \\\\
 $C_l$, $C_m$, $C_n$  & aerodynamic moment coefficients in body axes \\\\
 $C_x$, $C_y$, $C_z$  & aerodynamic force  coefficients in body axes \\\\
 $\bf F$              & non-viscous aerodynamic force  \\\\
 ${\bf F}^e$          & sum of the external forces \\\\
 ${\bf F}_v$          & viscous aerodynamic force  \\\\
 $\bf g$              & gravity acceleration \\\\
 $\bf H$              & $ ((H_{i j l}))$  third order apparent-mass tensor  \\\\
 $\bf J$              & inertia tensor \\\\
 $h$                  & height \\\\
 $h/c$                & dimensionless height \\\\
 $l$                  & vehicle length  \\\\
 $\bf L$              & transformation matrix from inertial to body frame  \\\\
 $m$                  & vehicle mass    \\\\
 $\bf M$              & $ ((M_{i j}))$ second order apparent-mass tensor \\\\
 $\bf n$              & local normal unit vector  \\\\
 $\hat{p}, \ \hat{q}, \ \hat{r}$  & $\ds \frac{p \ b}{V}, \  \frac{q \ c}{V}, \ \frac{r \ b}{V}$ dimensionless angular velocity components   \\\\
 ${\bf P}$            & linear momentum of the airstream \\\\
 ${\bf Q}$            & aerodynamic moment with respect to $c.g.$  \\\\
\end{tabular}
}

{
\no \begin{tabular}{ll}
 ${\bf Q}^e$          & sum of the external moment with respect to the $c.g.$  \\\\
 ${\bf r} \equiv (x, y, z)$ & aircraft c.g. position in Earth axes \\\\
 ${\bf r}_{c.g.}$     & aircraft c.g. position in Body axes \\\\
 ${\bf R}^{-1}$       & transformation matrix from $\bfomega$ to $\dot {\bfphi}$ \\\\
 $S$                  & wing area  \\\\
 $S_v$                & vehicle wetted surface \\\\
 $T$                  & kinetic energy of the stream flow  \\\\
 $\bf T$              & Thrust force \\\\
 $\bf v$              & $(u, \ v, \ w)$  inertial velocity in body axes \\\\
 $V$                  &  velocity modulus \\\\
 $W$                  & weight \\\\
 $x$, $y$, $z$        & Earth-fixed coordinates \\\\
 $x_B$, $y_B$, $z_B$  & body axes coordinates
\end{tabular}
}

\bigskip
\bigskip

\no {\sl Greek symbols }

{
\no \begin{tabular}{ll}
 $\alpha$, $\beta$    & angle of attack and sideslip, respectively \\\\
 ${\delta}_e$, ${\delta}_a$, ${\delta}_r$  &  elevator, ailerons and  rudder angles, respectively \\\\
 ${\delta}_T$         & $\ds \frac{\Pi}{\Pi_{max}}$  throttle level \\\\
 $\eta$               & propeller efficiency \\\\
 $\varphi$, $\vartheta$, $\psi$  & Euler angles \\\\
 $\phi$              & velocity potential  \\\\
 $\ds \frac{\partial \phi}{\partial {\bf v}}$ &
$(\ds \frac{\partial \phi}{\partial  u}, \ \frac{\partial \phi}{\partial  v}, \ \frac{\partial \phi}{\partial  w})$
  potential by unit velocity  \\\\
 $\gamma$            & flight path angle \\\\
\end{tabular}
}

{
\no \begin{tabular}{ll}
 $\rho$              & air density  \\\\
 $\tau$  & $ \equiv \sqrt{1+4 (h/c)^2} -2 h/c$  ground effect perturbative parameter \\\\
 $\bflambda$         & $(\lambda_x, \lambda_y, \lambda_z) \equiv {\bf L} \ (0, \  0, \ 1)^T$ ground normal unit vector in body axes \\\\
 $\bfomega$          & ($p$, $q$, $r$) angular velocity vector in body axes \\\\
 $\Pi$               & engine power \\\\
 $\zeta$             & turn rate
\end{tabular}
}

\bigskip
\bigskip

\no {\sl Subscripts }

{
\no \begin{tabular}{ll}
${\infty}$          &  value calculated out of ground effect \\\\
$c.g.$              & center of gravity \\\\
$a.c.$              & aerodynamic center \\\\
0                   & value calculated at $\varphi$ = $\vartheta$ = 0
\end{tabular}
}

\bigskip
\bigskip

\bc
 {\bf Introduction}
\ec

An important feature of ground effect is the
influence of the attitude on the vehicle dynamics.
This influence arises from the Euler angles,
which has an effect on the aerodynamic force
and moment coefficients. This is of paramount
importance for all aircraft maneuvers that
 are performed at very low altitudes, in particular
  in phases such as takeoff and landing.

The analysis of the aircraft stability and control
during take-off and landing requires an accurate knowledge
of the aerodynamic coefficients in ground effect$^1$.
Staufenbiel and Schlichting$^1$ analyzed the longitudinal
stability of an aircraft in ground effect. It was observed
that variations of the longitudinal stability caused
by the ground proximity is responsible
for substantial changes in the landing trajectories,
especially during the flare-out maneuvers.

More recentely the increased development of unmanned
aerial vehicles (U.A.V.) has resulted in the consideration
of these vehicles for missions at low altitudes$^2$.
Many U.A.Vs implement flight control programs to
accomplish the required mission profiles and are equipped
with plant control systems whose characteristics
take into account the forces and moments that are
 developed in ground effect$^{1, 2}$.

The vehicles which are considered in this study are
the "wing-in-ground-effect" (W.I.G.) crafts, also
called Ekranoplans, that are high-speed low-altitude
 flying vehicles that purposely utilize the
 favorable ground effect.
W.I.G. craft are of interest due to the peculiarities that they exhibit
in comparison to conventional airplanes.
Within the framework of high-speed low-altitude
transportation, the high aerodynamic efficiency
that these vehicles achieve together with low
wing aspect ratios and low structure weights make
them more suitable than conventional aircraft$^3$.

During horizontal flight with
$\varphi$ = $\beta$ = 0, the ground plane imposes a
boundary condition on the aerodynamic field that
symmetrically reduces the downward flow about the
aircraft. This results in a downwash reduction at
the tail together with an increase in the lift slope
of both wing and tail$^4$ that depends on the height
above the ground.
However if the vehicle flies with an arbitrary
orientation with respect to the inertial frame,
the ground alters the vehicle aerodynamics in such
a way that the pressure distribution on the
aircraft depends on both height and attitude.
Therefore the aerodynamic coefficients,
which are a function of both height and aerodynamic
angles, also vary with the Euler angles.
This effect, which modifies performance and
flying qualities, is observed by pilots when
 an aircraft is flown close to the ground.
For example, during flight tests on W.I.G.
vehicles$^{5}$, horizontal banked turns were
carried out to demonstrate that the sideslip,
rudder and ailerons angles are quite different
from those needed out of ground effect.
While these angles are quite small for steady
banked turns out of ground effect, for a W.I.G.
craft they are quite large.
Kornev and Matveev$^5$, considered the effect of
the roll angle and proposed semi-empirical
expressions for the aerodynamic coefficients
which are based on results from numerical
simulations.
They noted that, during a banked turn near the
ground, the vehicle develops  additional
aerodynamic forces and moments that depend
upon the roll angle.

From a safety perspective, a banked turn must be
performed with a limited roll angle and at a high
enough altitude to avoid the risk of touchdown.
Thus, a special maneuver strategy is adopted to
maintain the distance between the wing tip and
 ground surface.$^5$
This maneuver requires a very large turning radius
that strongly reduces the maneuverability of the
aircraft.
For this reason it is prudent to use a control
 system to minimize the turning radius$^6$.
The design of such controllers requires a knowledge
of the aerodynamic coefficients that are developed
in ground effect$^6$, while also accounting for
the influence of the attitude on aircraft's
forces and moments.

Another source of the significant variations of the
aerodynamic coefficients of a W.I.G. craft, is
nonzero sink rates.
The vertical velocity of the aircraft determines
the unsteady or dynamic ground effect which, in turn
produces aerodynamic forces and moments that depend
 upon the flight path angle.
Such an effect occurs during W.I.G. maneuvers as
 in the case of the vertical jumping$^5$, which
 is very short climbing phase that is used
 to avoid collisions with low obstacles.

In the present work the flight dynamics
of an aircraft in ground effect are examined through
an analytical procedure that estimates the
aerodynamic forces and moments and takes into
account the aforementioned effects, such as
attitude and sink rate.

In previous work Chang and Muirhead$^{7}$
experimentally examined the forces on low aspect
ratio wings with sink rate. Nuhait and Mook$^{8}$
numerically examined the phenomena using an unsteady
vortex-lattice method.
Nuhait and Zedan$^{9}$ analyzed the unsteadiness
induced by the vertical velocity using a vortex
lattice method, with a predetermined wake shape.
More recently Han and Yoon$^{10}$ investigated the
2D ground effect of flat plates in tandem
configuration using a discrete vortex method.
They observed that the unsteadiness has a marked
impact on the performance of a tandem configuration.

In Ref. 11, a mathematical model of flow about
lifting bodies, in steady and unsteady ground effect
is presented.
The linearized  equations of motion for the
longitudinal dynamics, in which the aerodynamic
coefficients are expressed by means of linear
derivatives with respect height, pitch angle and
 their time derivatives are developed.
Kumar$^{12}$, examined the dynamics of a W.I.G.
craft including the effect of a perturbation in the
forward velocity; the stability of the W.I.G. craft
was also examined from an eigenvalue analysis
of the linearized equations of motion.
Staufenbiel$^{13}$ analyzed the longitudinal
 stability of aircraft in ground effect using
 the characteristic equation. The nonlinear effects
were also examined by numerical calculation of
 the time-histories of the motion equations.

Although other references to the aerodynamic forces
 and moments in the ground effect$^{14-16}$ can be
  found, to the author's knowledge the effects
  exerted by the attitude of the aircraft on the
  flight dynamics have not been examined.
Therefore the objective of the present work is
to develop an accurate mathematical model that
may be used for the analysis of the performance
 and flying qualities of aircraft in ground effect.

In the present work the aerodynamic forces and
 moments are derived from the Lagrange equations.
 The Lagrangian function of the physical problem
 is given by the kinetic energy of the airstream,
which is expressed in terms of the aerodynamic and Euler angles.
This approach allows the forces and moments in
ground effect to be accurately formulated
and the variations in the performance and flying
 quality to be interpreted.

\bc
 {\bf  Equations of Motion}
\ec

In this section the rigid aircraft equations of motion, which are used to study
the vehicle dynamics in ground effect, are presented.
These equations are written in the form$^4$
\bea
 \begin{array}{l@{\hspace{1cm}}l}
m ( {\dot {\bf v}} +  \bfomega \times {\bf v} ) =  {\bf F}^e
\\\\
{\bf J} {\dot {\bfomega}} +  \bfomega \times {\bf J} {\bfomega} =  {\bf Q}^e
\\\\
\dot{\bf r} = {\bf L}^T {\bf v}
\end{array}
\label{eq_motion}
\eea
\bea
 \begin{array}{l@{\hspace{2.7cm}}l}
\dot{\bfphi } = {\bf R}^{-1} \bfomega
\end{array}
\label{eq_euler}
\eea
The elements of the diagonal matrix $\bf J$ are
\bea
J_{xx} = 0.02 \ m \ b^2, \ \ \ \ J_{yy} = 0.02 \ m \ l^2, \ \ \ \ J_{zz} = 0.02 \ m \ (b^2 + l^2)   \nonumber
\eea
${\bf F}^e = {\bf T} + m \ {\bf L}{\bf g} + {\bf F} + {\bf F}_v$ and  ${\bf Q}^e = {\bf Q}$
are the sum force and the moment with respect to center-of-gravity
 of the external forces, where $m \ {\bf L}{\bf g}$ represents
the weight force expressed in body axes.
$ \ {\bf F}_v$, which has a purely dissipative action,
 is the aerodynamic viscous force that is given as
$
{\bf F}_v = \ds - \frac{1}{2} \rho V^2 S {C_D}_f (\cos \alpha \cos \beta, \sin \beta, \sin \alpha \cos \beta)
$,
with ${C_D}_f = 0.047$ that does not depend on ground effect.
Thus ${\bf F}+ {\bf F}_v$ and ${\bf Q}$ are,  aerodynamic force and moment
which, in body axes are expressed as
\bea
{\bf F}+ {\bf F}_v \equiv
\left[\begin{array}{c}
X         \\\\
Y         \\\\
Z
\end{array}\right] =
\frac{1}{2} \rho V^2 S
\left[\begin{array}{c}
C_x         \\\\
C_y         \\\\
C_z
\end{array}\right]; \ \
{\bf Q} \equiv
\left[\begin{array}{c}
L         \\\\
M         \\\\
N
\end{array}\right] =
\frac{1}{2} \rho V^2 S
\left[\begin{array}{c}
b \ C_l         \\\\
c \ C_m         \\\\
b \ C_n
\end{array}\right]
\eea

The aircraft is assumed to be equipped by a propulsive system
which is a constant speed propeller driven by a reciprocating engine.
Thus, the thrust force $\bf T$, which is aligned with $x_B$,
is calculated by dividing the propulsive power
$\Pi_{max} \delta_T \eta$ by the velocity component $u$, where the
propeller efficiency $\eta$ is assumed to be constant and equal to 0.82.

\bc
 {\bf Analysis}
\ec

The goal of the present section is to examine how the attitude acts upon the flow-field about
a vehicle flying in ground effect.
Consider Fig. 1, a rigid aircraft which moves in a potential flow
and in close proximity to a rigid surface of infinite extent.
The attitude is represented by the vector
\be
{\bflambda} \equiv (- \sin \vartheta, \ \sin \varphi \cos \vartheta, \ \cos \varphi \cos \vartheta)
\label{lambda}
\ee
which is the normal unit vector of the ground surface in body frame.
The wakes behind the body are considered to be rigid and their shedding lines are assigned.
Therefore the velocity potential is the scalar product between
the derivative of the potential with respect to the velocity
and the flight velocity {\bf v} in body frame$^{17}$, i.e.
\be
\phi =  \ \ds \frac{\partial \phi}{\partial {\bf v} }   \cdot {\bf v}
\equiv \ds \frac{\partial \phi}{\partial {\bf v} } \cdot (\cos \alpha \cos \beta, \ \sin \beta, \ \sin \alpha \cos \beta ) \ V
\label{phi_v}
\ee
Since the velocity potential satisfies the Laplace equation and satisfies the integral equations derived
from the application of the Green's second identity, the derivative of the potential with respect to the
velocity is given by the sum of surface integrals$^{17}$. This derivative can be written in terms of
$\bflambda$ as
\be
\ds \frac{\partial \phi}{\partial {\bf v} } =
\frac{\partial \phi_0}{\partial {\bf v} } + \ds \frac{\partial^2 \phi_0}{\partial {\bf v} \partial {\bflambda} }
\ ({\bflambda}-{\bflambda}_0)
\label{phi_123}
\ee
where
$\ds \frac{\partial^2 \phi}{\partial {\bf v} \partial {\bflambda} }$
$\equiv (( \ds \frac{\partial^2 \phi}{\partial {v_i} \partial {\lambda}_j } ))$  is
the matrix of the second order derivatives of the potential.

While Eq.(\ref{phi_v}) yields the relationship between $\phi$ and the aerodynamic angles,
the second term of Eq. (\ref{phi_123}) accounts for the effect of the attitude on the vehicle aerodynamics.
The kinetic energy of the airstream using Eqs. (\ref{phi_v}) and (\ref{phi_123}), is
\be
T = \frac{1}{2} \rho \int \int_{S_v} {\bf v} \cdot {\bf n} \  {\phi} \  dS
= \frac{1}{2} {\bf v} \cdot {\bf M} \ {\bf v}
\ds + \ \frac{1}{2} {\bf v} \cdot \{  {\bf H} ( {\bflambda}-{\bflambda}_0 )   \} \ {\bf v}
\label{T}
\ee
where
\be
{\bf M} = \ds \int \int_{S_v} \ds \frac{\partial {\phi}_0}{\partial \bf v } \otimes {\bf n} \ dS,
\ \ {\bf H} = \ds \int \int_{S_v} \ds \frac{\partial^2 {\phi}_0}{\partial {\bf v} \partial {\bflambda} } \otimes {\bf n} \ dS
\label{MG}
\ee
are second and the third order tensors.
Eq. (\ref{T}),  describes the kinetic energy of the stream in terms of aerodynamic angles;
the effect of the vehicle attitude is accounted for through the second term,
where the quantity  ${\bf H} (\bflambda- {\bflambda}_0)$ is a second order tensor that is calculated as
$
{\bf H} ({\bflambda}- {\bflambda}_0 ) \equiv H_{i j k} (\lambda_k - {\lambda_k}_0)
$.

It is worthwithe to highlight some characteristics of the kinetic energy.
First, for flat ground, the heading angle does not modify the
aerodynamic field about the aircraft, therefore in Eq. (\ref{T}) the
kinetic energy does not depend on $\psi$.
Furthermore, because of the vehicle symmetry about the plane ($x_B$, $z_B$), $T$ is an even
function of $\varphi$ and $\beta$.
Therefore both ${\bf M}$ and ${\bf H}$ are expressed in the form
\bea
{\bf M} =
\left[\begin{array}{ccc}
\ds  M_{1 1} &      0     & M_{1 3} \\\\
\ds  0         &  M_{2 2} & 0         \\\\
\ds  M_{1 3} &      0     & M_{3 3}
\end{array}\right]; \ \
{\bf H}  =
\left[\begin{array}{ccc}
\ds  {\bf h}_{1 1} &      0          & {\bf h}_{1 3}\\\\
\ds  0             &  {\bf h}_{2 2}  & 0         \\\\
\ds  {\bf h}_{1 3} &      0          & {\bf h}_{3 3}
\end{array}\right]; \ \
\label{9a}
\eea
with ${\bf h}_{i j} \equiv (h_{i j 1}, \ 0, \ h_{i j 3} )$, $ \ i, j \ne 2$.

Eqs. (\ref{MG}) establish that $\bf M$ and $\bf H$ are functions of $h/c$ that do not vary with the Euler and aerodynamic angles.
In the present work, both the tensors $\bf M$ and $\bf H$ are expressed in terms of the dimensionless height through the parameter $\tau$ $^{11}$
\be
f = f_{\infty} + f_1 \tau + f_2 {\tau}^2
\label{tau_e}
\ee
where $\tau$ is defined in the nomenclature.

\bc
 {\bf Calculation Method}
\ec

In the equations of motion, the representation of the aerodynamic force and moment coefficients in terms  of the Euler angles, flight path angle and dimensionless height, is necessary to interpret the performance and flying qualities in ground effect.
This section deals with the calculation of aerodynamic actions developed by a vehicle in ground proximity.
The procedure presented here allows the calculation of the aerodynamic forces and
moments through the Lagrange equation method.
This approach has been already applied to the calculation of the aerodynamic forces and moments of ultralight aircraft that are flown in the presence of a wind gradient$^{18}$.
To derive the expression of force and moment, the kinetic energy of the potential flow that is generated by the vehicle is considered.
Then the aerodynamic force ${\bf F}$ and moment ${\bf Q}$ are calculated using the Lagrange equation method in the general form$^{17, 18}$
\be
{\bf F} =
- \ds \frac{d}{dt} \frac {\partial T} {\partial {\bf v}}
-{\bfomega} \times \frac {\partial T} {\partial {\bf v}}
\label{F_1}
\ee
\be
{\bf Q} = {\bf Q}_{a.c.} + ({\bf r}_{a. c.} - {\bf r}_{c. g.}) \times {\bf F} \equiv
 \ds -{\bf v} \times \frac {\partial T} {\partial {\bf v}} +({\bf r}_{a. c.} - {\bf r}_{c. g.}) \times {\bf F}
\label{Q_1}
\ee
Substituting into Eqs. (\ref{F_1}) and (\ref{Q_1}) the expression of $T$ that is  given by Eq. (\ref{T}),
the aerodynamic force and moment are
\bea
\begin{array}{l@{\hspace{1cm}}l}
{\bf F} = \ds -  [ \ \dot{\bf M}  + \dot{\bf H}({\bflambda}-{\bflambda}_0)] {\bf v} -   {\bf H} \dot{\bflambda}   {\bf v}
 - \ds  {\bfomega} \times  [ \ {\bf M}  +  {\bf H} ({\bflambda}- {\bflambda}_0  )  \ ] {\bf v}
\ds
\end{array}
\label{F_2}
\eea
\bea
\begin{array}{l@{\hspace{1cm}}l}
{\bf Q} = - \ds {\bf v} \times \ds  {\bf M}  {\bf v} \  - \ {\bf v} \times [{\bf H} ({\bflambda}- {\bflambda}_0)] {\bf v} \
+ \ ({\bf r}_{a. c.} - {\bf r}_{c. g.}) \times {\bf F}
\end{array}
\label{Q_2}
\eea

Eq. (\ref{F_2}) states that ${\bf F}$ is the sum of three terms, the first of which
contains the two quantities $\dot{\bf M}$ and $\dot{\bf H}$ that are
\bea
 \begin{array}{l@{\hspace{1cm}}l}
\dot{\bf M} =  \ds \frac{\partial {\bf M}}{\partial {\bf r}_w} \ {\bf v} + \ds \frac{\partial {\bf M}}{\partial h} \ V \sin \gamma
 \equiv \ds  \frac{\partial M_{i j}}{\partial  {r_k}_w} \ v_k + \frac{\partial M_{i j}}{\partial h} \ V \sin \gamma  ;  \\\\
\dot{\bf H} = \ds \frac{\partial {\bf H}}{\partial {\bf r}_w} \ {\bf v} + \ds \frac{\partial {\bf H}}{\partial h} \ V \sin \gamma
 \equiv  \ds \frac{\partial H_{i j k}}{\partial  {r_l}_w} \ v_l  + \frac{\partial H_{i j k}}{\partial h} \ V \sin \gamma
\end{array}
\label{M_r}
\eea
According to the classical lifting bodies theory, $\dot{\bf M}$ and $\dot{\bf H}$  are proportional to the velocity circulation around the aircraft. They represent the time derivatives of ${\bf M}$ and ${\bf H}$  which are related to the shedding wake surfaces past the aircraft and to the sink rate.
$\ds \frac{\partial {\bf M}}{\partial {\bf r}_w}$ and $\ds \frac{\partial {\bf H}}{\partial {\bf r}_w}$
are the apparent mass terms relative to the unit of wake length ${\bf r}_w$ which gives the aerodynamic force developed in horizontal flight, whereas both $\ds \frac{\partial {\bf M}}{\partial h}$ and $\ds \frac{\partial {\bf H}}{\partial h}$ produce the unsteady ground effect when $\gamma \ne $ 0.
This last result is a consequence of the  observation that the pressure on the vehicle surface depends on the sink rate through the  Bernoulli theorem
$
p \ds = \rho (const - \frac{{\bfv} \cdot {\bfv}}{2} - \frac{\partial \phi}{\partial t})
\label{Berny}
$,
where $\bfv$ is the local velocity of the stream, while
$\ds \frac{\partial \phi}{\partial t} = \frac{\partial \phi}{\partial h} \ V \sin \gamma $
is  the unsteady term caused by the sink rate.
The second term in Eq. (\ref{F_2}) depends on $\dot{\bflambda}$, and represents a force that varies with the angular velocity.
In fact it can be written taking into account that $\dot{\bflambda}$ is related to the angular velocity through  Eq.(\ref{eq_euler})
\be
\dot{\bflambda}= \ds \frac{\partial \bflambda}{\partial {\bfphi}} \ \dot{\bfphi} = \ds \frac{\partial \bflambda}{\partial
 {\bfphi}} \ {\bf R}^{-1} {\bfomega}
\ee
where
\bea
{\frac{\partial \bflambda}{\partial {\bfphi}} }=
\left[\begin{array}{ccc}
\ds \frac{\partial \lambda_x}{\partial {\varphi}} &  \ds \frac{\partial \lambda_x}{\partial {\vartheta}}  & \ds \frac{\partial
 \lambda_x}{\partial {\psi}} \\\\
\ds \frac{\partial \lambda_y}{\partial {\varphi}} &  \ds \frac{\partial \lambda_y}{\partial {\vartheta}}  & \ds \frac{\partial
 \lambda_y}{\partial {\psi}} \\\\
\ds \frac{\partial \lambda_z}{\partial {\varphi}} &  \ds \frac{\partial \lambda_z}{\partial {\vartheta}}  & \ds \frac{\partial
 \lambda_z}{\partial {\psi}}
\end{array}\right] \ \
\equiv
\left[\begin{array}{ccc}
\ds  0                             & -\cos \vartheta                & 0 \\\\
\ds   \cos \varphi \cos \vartheta  & -\sin \varphi \sin \vartheta   & 0 \\\\
\ds  -\sin \varphi \cos \vartheta  & -\cos \varphi \sin \vartheta   & 0
\end{array}\right] \ \
\label{lambda_phi}
\eea
Hence, this is a contribution to the rotational derivatives that tend to zero as $h/c \rightarrow \infty$.
Also the last term of Eq. (\ref{F_2}) gives a contribution to the rotational derivatives, which in turn depends upon the height and vehicle attitude.

In Eq. (\ref{Q_2}) for the moment, the first term is a function of $\alpha$ and $\beta$,
whereas the second term is a function of the attitude variations.

Thus, force and moment are
\bea
\begin{array}{l@{\hspace{1cm}}l}
{\bf F} = \ds -  \{ \ds \frac{\partial {\bf M}}{\partial h} \ V \sin \gamma +
 \frac{\partial {\bf M}}{\partial {\bf r}_w} \ {\bf v}  +
 \ds [ \frac{\partial {\bf H}}{\partial h} V \sin \gamma +
  \frac{\partial {\bf H}}{\partial {\bf r}_w} \ {\bf v} ]  ({\bflambda}-{\bflambda}_0) \} {\bf v} \\\\
-  \{ {\bf H}  \  \ds \frac{\partial \bflambda}{\partial {\bfphi}} \ {\bf R}^{-1} {\bfomega} \}  {\bf v}
- \ds  {\bfomega} \times  \{ \ {\bf M}  +  {\bf H} ( {\bflambda}- {\bflambda}_0  )  \} {\bf v}
\ds
\end{array}
\label{F_3}
\eea
\bea
\begin{array}{l@{\hspace{1cm}}l}
{\bf Q} = - \ds {\bf v} \times \ds  \{ {\bf M}   + {\bf H} ({\bflambda}- {\bflambda}_0)  \} \  {\bf v}
+({\bf r}_{a. c.} - {\bf r}_{c. g.}) \times {\bf F}
\end{array}
\label{Q_3}
\eea
Therefore, Eqs. (\ref{F_3}) and (\ref{Q_3}) allow the determination of $\bf F$ and $\bf Q$
in ground proximity and account for the attitude and sink rate.

\bc
 {\bf  Validation of the Method}
\ec

To validate the method, several comparisons with existing data in the literature are presented.

The  aerodynamic force and moment coefficients are evaluated from Eqs. (\ref{F_2}) and (\ref{Q_2}),
 where  ${\bf M}$ and  ${\bf H}$ are numerically calculated using Eq. (\ref{MG}).
The velocity potential is computed by an unsteady vortex-lattice code, developed by the author,
that takes into account the ground effect using the method of images.
Specifically the presence of the ground is simulated by placing a specular image of the
 lifting body at an equal distance below the ground plane.
Thus, two symmetrically positioned lifting bodies are considered to determine a flat ground surface.

The results of the first case are presented in Fig. 2a.
The figure shows increments of the lift coefficient in terms of $h/c$ for a delta wing
 having an aspect ratio equal to $1.456$; $h$ is measured at the midpoint of the root chord.
The simulation is computed for an angle of attack of $22.1^\circ$ and a flight path angle equal to $-10^\circ$.
The dashed line represents the data of Ref. 8, while the symbols are the experimental data of Chang and Muirhead$^{7}$.
The figure shows that the lift coefficients obtained by the present method, shown by the solid line, are in good agreement with the those of both the Refs. 7 and  8.

Figs. 2b, 2c, and 2d show, respectively, the variations of lift, drag and pitching moment coefficients for a delta wing having an aspect ratio of $1.456$, in terms of the dimensionless height, measured at the trailing edge.
The simulations are made for $\alpha$ = 10$^{\circ}$ and $\gamma$ = $0$,  $-20^\circ$.
The difference between the values calculated by the present method and those reported in Ref. 8 is always less than 6 \%.

The next case, shown in Figs. 2e, 2f and 2g shows the effect of aspect ratio on the aerodynamic coefficients. The plots show three different calculations for $AR$ = 1.456, 1.072 and 0.705. According to Ref. 8 the height is   measured at the trailing edge and the solid symbols represent the data given in Ref. 8.
The calculations show the unsteady ground effect, that is obtained for $\alpha$ = 10$^\circ$ and $\gamma$ = -20$^\circ$.
The results are in good agreement with those of Ref. 8.

These comparisons demonstrate that the present method provides an accurate estimation of the aerodynamic coefficients for the examined wings in steady and unsteady ground effect over a wide range of $h/c$ variations.

\bc
 {\bf  Reference Vehicle}
\ec

Consider a model of a vehicle sketched in Fig. 3, the main parameters of which are reported in Table 1.
The vehicle is chosen according to the "Lippisch Design"$^{19}$ that combines a high
positioned tail with an inverted delta wing having negative dihedral along the leading edge.
 Such configuration is longitudinally stable at different $h/c$, as a result of an adequate
 aerodynamic moment developed by the tail which is located out of ground effect.$^{19}$

The aerodynamic coefficients of the aircraft are estimated using Eqs. (\ref{F_3}) and (\ref{Q_3}),
where the matrices ${\bf M}$ and  ${\bf H}$ are numerically evaluated by means of Eqs. (\ref{MG}).

The velocity potential is calculated by the boundary element code VSAERO$^{20}$
 that is capable of solving the complex aerodynamic field around an aircraft
 in the presence of the ground.

The incremental aerodynamic coefficients  caused by the control angles
are assumed to be constant in any situation. They are
\vspace{-0. mm}
\bea
 \begin{array}{l@{\hspace{-2.cm}}l}
\Delta C_x =  -0.029  \ {\delta_e}^2  -0.0493 \ {\delta_r}^2  - 0.113  \ {\delta_a}^2 \\\\
\Delta C_y =    0.157 \ {\delta_r}    \\\\
\Delta C_z =  -0.32 \ {\delta_e} -0.0112  \ {\delta_r}^2 + 0.1352 \ {\delta_a}^2  \\\\
\Delta C_l =   -0.07  \ {\delta_a} + 0.04  \ {\delta_r}       \\\\
\Delta C_m =  -0.923  \ {\delta_e} + 0.20  \ {\delta_a}^2 - 0.0055 \ {\delta_r}^2  \\\\
\Delta C_n =  +0.0035 \ {\delta_a} -0.072  \ {\delta_r}
\end{array}
\label{27}
\eea
where the control angles are expressed in radian.

The results that deal with the vehicle characteristics in ground effect are next presented.
In the subsequent figures the lines represent the calculation made using Eq. (\ref{F_3})
and (\ref{Q_3}), whereas
the symbols indicate the aerodynamic coefficients derived from surface integrals of the
pressure forces computed in the code.

Fig. 4 shows how the aerodynamic characteristics vary with the height.
The lift coefficient and aerodynamic efficiency in steady ground effect are shown
in terms of the angle of attack for $h/c$ = 0.167, 0.614, 2.792, $\infty$, with $\varphi$ = $\beta$ = 0.
The variations in $C_L$ are similar to the corresponding data presented in Ref. 1, while $C_L/C_D$, whose maximum value varies from about 7 (out of ground effect) to
more than 12 (h/c = 0.167), is in good qualitative agreement with the data presented in Ref. 11.

The influence of the pitch angle on the aerodynamic coefficients in steady ground effect is shown
in Fig. 5 for $C_L$, $C_D$ and $C_m$ vs. $\vartheta$.
The calculations are carried out at $h/c$ =1, with  $\gamma = \beta  =  \varphi  =  0$, for the two cases ${\bf H}$ = 0 (dashed lines) and ${\bf H} \ne$ 0 (solid lines).
The case with ${\bf H}$ = 0 corresponds to the usual assumption of the aerodynamic coefficients that
do not depend on $\varphi$ and $\vartheta$, whereas the calculation with ${\bf H} \ne$ 0 accounts
for the attitude effects.
It is worthwhite to emphasize that the pitch angle produces marked
variations on all the aerodynamic coefficients.
In the case ${\bf H} \ne$ 0, $C_L$ and $C_D$ exhibit significant increments when $\vartheta >$ 0, while the attitude acts upon the pitching moment so as to result in a quasi-linear variation of $C_m (\vartheta)$.
For each coefficient, the maximum percentage differences between the two cases can reach $20 \%$
of the corresponding value calculated at $\varphi$ = $\vartheta$ = 0.

The effects of the lateral attitude are next examined.
Fig. 6 shows all the aerodynamic coefficients in body axis in terms of roll angle, calculated for $\alpha = 5^{\circ}$ and
$\beta = \gamma = 0$.
Because of the vehicle symmetry, the longitudinal coefficients, $C_x$, $C_z$ and $C_m$
and the lateral ones, $C_y$, $C_l$ and $C_n$ are, respectively, even and odd functions of $\varphi$.
For $\vert \varphi \vert <10^{\circ}$, the increment in the longitudinal coefficients
can be greater than $15 \%$ for $h/c < 1$.
Note that the variations of $C_x$ and $C_z$ correspond to increases of  $C_L/C_D$
that can lead to an improvement in the vehicle performance during turning maneuvers.
As far as the lateral coefficients are concerned, linear variations are observed, for $h/c < 1$
that can be greater than those caused by a sideslip angle $\vert \beta \vert$ = $\vert \varphi \vert$.
Therefore, the attitude effects modify the force and moment coefficients;
the degree of change depends upon the dimensionless height.

The influence of the flight path angle on the aerodynamic coefficients is next examined.
Fig. 7 shows $C_L$, $C_D$ and $C_m$ in terms of  $h/c$,  for  $\gamma$= -5$^o$, 0, +5$^o$.
In these results, for which $\alpha = 5^\circ$, $\beta = \varphi = 0$, it is evident that
 the ground effect is stronger for negative flight path angles at relatively low $h/c$.

The proposed model also includes the effects of angular velocity.
Force and moment rotational derivatives are calculated from the derivative of
 Eqs. (\ref{F_3}) and (\ref{Q_3}) with respect to $p$, $q$ and $r$.
Fig. 8 shows the dimensionless  longitudinal rotational derivative
in terms of the angle of attack, calculated at $ \beta = \varphi =0$, for various $h/c$.
Large variations in the derivative can be seen.

Comparisons between the present results and those reported in Kornev$^5$
 and Staufenbiel$^{13}$ are next presented.
In Ref. 5, the rotational derivative ${C_L}_q$ of the W.I.G. craft ELA01 is given.
Figure 9 shows the comparison between the derivatives obtained in Ref. 5 (horizontal axis)
and these calculated using the proposed method (vertical axis).
In Fig. 9a the derivative ${C_L}_q$ calculated at
$h/c = 0.1$ is represented, whereas Fig. 9b reports ${C_L}_{\gamma}$ and ${C_m}_{\gamma}$.
Although the vehicle is geometrically different with respect to the reference vehicle, the present results are in relatively good agreement with the data of Ref. 5.
As for Ref. 13, the data concerns the W.I.G. craft X-113,
which is also designed according to the Lippish criteria.
Although the two vehicles have some geometrical differences,
Table 2 shows that the aerodynamic coefficients calculated by the present method
are in good agreement with those from Ref. 13.

\bc
{\bf  Results and Discussions}
\ec

In what follows, to study the influence of the ground effect on state and
control variables at trim, some significant situations  which correspond to the level and turning flight will be considered. Furthermore the vehicle stability will be investigated by means of the eigenvalues analysis applied to the linearized motion equations, whereas the nonlinear analysis of the vehicle motion will be made integrating the full set of the equations of motion.

Eqs. (\ref{eq_motion}) and (\ref{eq_euler}) are used as equations of motion and the contribution of the three control angles to the aerodynamic forces and moments is taken into account through incremental force and moment coefficients as reported in Eqs. (\ref{27}).

As a first result, Fig. 10 summarizes the trim condition in level flight for $\beta$ = 0 at different $h/c$.
The diagrams show the angle of attack, elevator angle and required throttle in terms of the flight speed.

Since in ground effect the lift coefficient increases when $h/c$ diminishes,
the angle of attack decreases with $h/c$ (Fig. 10a).
$\delta_e$ (Fig. 10b) monotonically increases with $V$, which indicates that there is a
stable behavior except the lower velocities and altitudes, where the curves show decreasing
branches ($h/c \ < 1$).
This last feature only occurs at high angle of attack and is the result of the simultaneous
effects of $\vartheta$ and $\alpha$ on the aerodynamic forces and pitching moment.
Fig. 10c shows the required throttle as a function of the flight speed. Each curve exhibits
a distinct minimum value that increases with $h/c$.
Since the ground effect is stronger at high angle of attack, the throttle for $h/c$= 0.5 and 1
 is less than the throttle for $h/c \rightarrow \infty$ only for $V  <  55 \ m \ s^{-1}$.

In order to evaluate the attitude effect on symmetric flight situations, two different
trim conditions in straight flight, at $h/c = 1$, are examined (Fig. 11).
The continuous and dashed lines indicate the solutions calculated for ${\bf H} \ne 0$ and ${\bf H }= 0$, respectively.
The two solutions show substantial divergences which are caused by the different aerodynamic coefficients
 in the two cases.
At lower flight speeds the angle of attack shows important differences between the two solutions (Fig. 11a), whereas minor divergences are evident in Fig. 11b, where the elevator angle is displayed.
The little angles of attack obtained for ${\bf H} \ne $ 0 are in agreement with the data given in Fig. 5
that shows relatively high lift coefficients when the attitude effects are accounted for.
As a result, for ${\bf H} \ne $ 0, lower flight speeds than those calculated for ${\bf H} =$ 0, are obtained.
Also the required throttle (Fig. 11 c) depends upon the attitude.  In fact, $\vartheta$ acts on the aerodynamic forces in such a way that the throttle exhibits significant reductions for velocities less than 40 m s$^{-1}$.

The third case, shown in Fig. 12,  analyzes the effect of the roll angle on a banked maneuver.
The aircraft carries out a "truly banked" turn in the horizontal plane, and
the angular velocity $(p, \ q, \ r)$ $\equiv$
$(-\sin \vartheta,$ $ \cos \vartheta \sin \varphi,$ $\cos \vartheta \cos \varphi)$ $\zeta$ is vertical with $\zeta = 5^\circ s^{-1}$.
In this maneuver the sum of weight and centrifugal force at the aircraft c.g., is in the vehicle symmetry plane $(x_B, z_B)$.
In order to avoid possible touchdown during the maneuver, it is assumed that $h/c = 2$ in all situations.
In the diagrams the solutions which correspond to ${\bf H} =0$ and ${\bf H} \ne 0$ are shown.
Fig. 12a shows that there are higher differences in the angles of attack between the two cases
at lower flight velocities.
As in the case of a "truly banked" maneuver that is performed out of ground effect, the solution for
 ${\bf H} =0$ shows very small sideslip angles (Fig. 12b), whereas for ${\bf H} \ne 0$ nonzero sideslip are observed.
The reason of such discrepancy is due to $\varphi$ that causes additional force and moment terms
that are thus balanced by nonzero sideslip angles.
The roll angle (Fig. 12c) exhibits little differences between the two solutions over the entire range of the flight speed, whereas more significant differences are seen in the pitch angle (Fig. 12d).
If ${\bf H} = 0$, ailerons and rudder are very little as in a banked turn at high altitude,
whereas, for ${\bf H} \ne 0$ the attitude effects lead to a disagreement between the two solutions.
Fig. 12h shows the effect on throttle for both the cases.
As a result of the aerodynamic efficiency increments caused by the nonzero roll angles (see Fig. 6),
lower throttle levels are needed for ${\bf H} \ne 0$.
This means that, at least in these flight conditions, the drag increment due to the maneuver is in part counterbalanced by the drag reduction caused by $\varphi$.

The vehicle stability is next evaluated through the eigenvalue analysis applied to the linearized equations of motion, where the various modes are identified by means of the eigenvector analysis.
The root locus is calculated for straight flight at the speed of 40 m/s, varying $h/c$ from $\infty$ to about 1.

The attitude effects on the vehicle stability are summarized in the two sets of Figs. 13a, 13b and 13c, 13d which show, respectively, the two cases ${\bf H} =0$ and ${\bf H} \ne 0$.
Because of the different aerodynamic force and moment coefficients developed in the two situations,
the eigenvalues vary with $h/c$ in a different manner.
However, out of ground effect, the vehicle shows three oscillating stable modes (which are phugoid, short period and Dutch roll) and two aperiodic modes (that are the roll convergence (stable) and the spiral mode (unstable)).

The case ${\bf H} =0$ is first discussed.
Phugoid is a stable mode in the entire range of $h/c$, whose eigenvalues significantly vary for $h/c  < 4$ (see Fig. 13a and 13b). Its imaginary and real parts, respectively, increase and diminish as  $h/c\rightarrow$ 0.
At each $h/c$ the short period is a stable mode that has significant variations in the eigenvalues
for $h/c < $ 4 that are due to the nonlinearities of $C_m (\vartheta)$ (see Fig. 5).
The Dutch roll eigenvalues show smaller variations. Their imaginary part do not change with $h/c$, while the real part displays more marked variations.
As for the non oscillating modes, the positive eigenvalue associated with the spiral mode increases as $h/c \rightarrow $ 0.
Although the rolling convergence remains a stable mode, it presents a negative eigenvalue that rises as soon as $h/c$ diminishes.

Figs. 13c and 13d show the root locus for ${\bf H} \ne 0$.
The attitude effect produces significant differences with respect to the previous case.
The phugoid mode exhibits the same behavior with the exception of lower $h/c$, where
it becomes unstable (See Fig. 13c). Such instability is produced by the combined effect of both pitch angle and height upon the lift coefficient which is more pronounced at lower $h/c$.
With respect to the case with $\bf H$ =0, the attitude influences the pitching moment such that the derivative ${C_m}_{\alpha}$ is almost constant with the height; therefore the short period eigenvalues have smaller variations with $h/c$.

Pronounced discrepancies between the two cases are evident in terms of the lateral modes.
It is seen that the Dutch roll eigenvalues are quite different from those of the preceding case.
They show increments for $h/c < $ 4 that make the mode unstable at $h/c \simeq$ 1.5.
More important divergence are apparent for spiral mode and rolling convergence.
The Spiral mode is completely changed.
It  becomes stable at about $h/c =$ 20, whereas, for
$h/c \simeq 10.5$, it  degenerates, together with the rolling convergence, in a new mode called oscillating rolling that is the consequence of the roll angle influence on the aerodynamics force and moment.
The oscillating rolling, which occurs under a certain  $h/c$, is a stable oscillating mode that modifies the lateral stability.
Also the rolling convergence depicts nonnegligible variations. Its eigenvalue strongly varies for $11 < h/c < 20$, while, as seen, for $h/c \simeq 10.5$ it disappears.

In the reported calculations the attitude considerably changes the lateral stability of the vehicle.
The stabilization of the spiral mode, the disappearance of both spiral and rolling convergence and
the subsequent genesis of the oscillating rolling yield a rather different lateral stability than one where the attitude is not accounted for.

In order to estimate the dynamic response, time histories, which are calculated by
integrating the full set of the motion equations, for ${\bf H} \ne $ 0 are next examined.
The first simulation takes place in the vertical plane at various initial heights $h_0$, where the vehicle, initially in level flight at equilibrium at the speed of 40 m s$^{-1}$, is perturbed by a pitch rate of 0.8 s$^{-1}$.
Fig. 14 gives the time histories of the perturbed height and angle of attack. In the diagrams phugoid and short period modes can be seen. Note that the characteristic times of both the modes are in agreement with the  previously calculated eigenvalues. In all the cases the phugoid mode is persistent during the entire simulation period, whereas the short period, due to its strong damping, only acts at the beginning of the motion. As seen in
Fig. 13, as soon as $h_0/c$ diminishes,  the phugoid period decreases and its amplitude is more damped.

The second case, shown in Fig. 15, concerns a perturbed lateral motion, where the initial condition is the same
 one of the previous case, while the perturbed action is now given by a roll rate of 0.2 s$^{-1}$.
Out of ground effect ( $h_0/c \rightarrow \infty$), it is possible to identify rolling convergence,
 Dutch roll and spiral mode.
For each initial height, the time-histories show characteristic times that are in agreement with the data obtained through the eigenvalues analysis.
For $h_0/c$ = 4, the oscillating rolling mode is apparent together with the Dutch roll, while both rolling convergence and spiral modes disappear.
It is interesting to note that, at $h_0/c$ = 2, the Dutch roll is obscured by the oscillating rolling which produces the main effect. This result can be also obtained by analyzing the eigenvectors of each mode.

\bc
 {\bf  Conclusions}
\ec

In this study the dynamics in ground effect is investigated accounting for
the influence of the attitude on the vehicle dynamics.
The Lagrange equations method applied to lifting bodies is used to express the forces and moments
that are  developed in ground effect.
This method has general validity and contains the important elements for describing ground effect.
In particular the method leads to the determination of aerodynamic force and moment coefficients
in terms of roll, pitch and flight path angles.
Comparisons between existing data in the literature and those obtained with the present method show that the procedure provides accurate results for aerodynamic force and moment
of wings in steady and unsteady ground effect.
Once the method is validated, the aerodynamic coefficients of a vehicle geometry are examined.
The attitude effects on the vehicle aerodynamics are identified and
the unsteadiness induced by nonzero sink rates are explained.
The trim calculation for the banked turn clarifies the behavior observed in ground effect, while the stability analysis yields explanations about the flying qualities with particular reference to the spiral mode stabilization and to the existence of the oscillating rolling.
The nonlinear analysis, that is obtained by integrating the full set of the equations of motion,
gives results in accordance with those obtained by means of the linear stability.

\bc
{\bf Acknowledgment}
\ec

This work is partially supported by Italian Ministry of University (MIUR).

\bc
{\bf References}
\ec

\no \hspace*{-1.50mm}$^1$Staufenibel  R.W. and Schlichting, U.J., ``Stability of Airplanes in Ground Effect'',
           {\it Journal of Aircraft}, Vol. 25, No. 4, 1988, pp. 289-294.

\medskip

\no \hspace*{-1.50mm}$^2$Walsh, D. and Cycon J.P., ``The Sikorsky Cypher UAV: A Multi-purpose
 Platform with
Demonstrated Missions Flexibility'', Proceedings of the Annual Helicopter Society 54th Annual
 Forum,
Washington DC, May 1998, pp. 1410-1418.

\medskip

\no \hspace*{-1.50mm}$^3$Lange  R.H and Moore, J.W., ``Large Wing-in-Ground Effect Transport Aircraft'',
           {\it Journal of Aircraft}, Vol. 17, No. 4, 1979, pp. 260-266.

\medskip

\no \hspace*{-1.50mm}$^4$Etkin, B., {\it  Dynamics of Atmospheric Flight.}
           John Wiley \& Sons, New York, 1972,  pp.104-152, 259-260.

\medskip

\no \hspace*{-1.50mm}$^{5}$Kornev N. and Matveev K. ``Complex Numerical Modeling of Dynamics a Crashes of Wing-in-Ground Vehicle'', AIAA 2003-600, presented at 41st Aerospace Sciences Meeting and Exhibit 6-9 January 2003, Reno Nevada.

\medskip

\no \hspace*{-1.50mm}$^{6}$Yang, Hee Joon, {\it  A Study on the Enhancement of Lateral Motion of a Wing-in-Ground Effect Ship},
           MS Thesis, Department of Naval Architecture and Ocean Engineering,Seoul National University Korea, 1999.

\medskip

\no \hspace*{-1.50mm}$^{7}$Chung Chang R. and Muirhead, V.U., ``Effect of Sink Rate on Ground Effect of Low-Aspect-Ratio Wings'',
           {\it Journal of Aircraft}, Vol. 24, No. 3, 1987, pp. 176-180.

\medskip

\no \hspace*{-1.50mm}$^{8}$Nuhait, A.O and Mook, D.T., ``Numerical Simulation of Wings in Steady and Unsteady Ground Effects'',
           {\it Journal of Aircraft}, Vol. 26, No. 12, 1989, pp. 1081-1089.

\medskip

\no \hspace*{-1.50mm}$^{9}$Nuhait, A.O and Zedan, M.F., ``Numerical Simulation of Unsteady Flow Induced by a Flat Plate Moving
Near Ground'',
           {\it Journal of Aircraft}, Vol. 30, No. 5, 1993, pp. 611-617.

\no \hspace*{-1.50mm}$^{10}$Han C., Yoon Y. and Cho J., ``Unsteady Aerodynamic Analysis of Tandem Flat Plates in Ground Effect'',
           {\it Journal of Aircraft}, Vol. 39, No. 6, 2002, pp. 1028-1034.

\medskip

\no \hspace*{-1.50mm}$^{11}$Rozhdestvensky, K.V., {\it  Aerodynamics of a Lifting System in Extreme Ground Effect}, Springer-Verlag, 2000,  pp.263-318.

\medskip

\no \hspace*{-1.50mm}$^{12}$Kumar, P.E., "Some Stability Problems of Ground Effect Vehicle in Forward Motion",  {\it Aeronautical Quarterly}, Vol. 18, Feb, 1972,  pp.41-52.

\medskip

\no \hspace*{-1.50mm}$^{13}$Staufenbiel R., ``Some Nonlinear Effect in Stability and Control of Wing-in-Ground Effect Vehicles'',
           {\it Journal of Aircraft}, Vol. 15, No. 8, 1978, pp. 541-544.

\medskip

\no \hspace*{-1.50mm}$^{14}$Tuck, E.O., ``Nonlinear Extreme Ground Effect on Thin Wings of Arbitrary Aspect Ratio'',
           {\it Journal of Fluid Mechanics}, Vol. 136, 1983, pp. 73-84.

\medskip

\no \hspace*{-1.50mm}$^{15}$Tuck, E.O., ``Nonlinear Unsteady One-Dimensional Theory for Wing in  Extreme Ground Effect'',
           {\it Journal of Fluid Mechanics}, Vol. 98, 1980, pp. 33-47.

\medskip

\no \hspace*{-1.50mm}$^{16}$Newman, J.N., ``Analysis of Small-Aspect-Ratio Lifting Surfaces in Ground Effect'',
           {\it Journal of Fluid Mechanics}, Vol. 117, 1982, pp. 305-314.

\medskip

\no \hspace*{-1.50mm}$^{17}$Lamb, H., ``On the Motion of Solids Through a Liquid'', {\it Hydrodynamics},
                6th ed., Dover, New York, 1945, pp. 160-201.

\medskip

\no \hspace*{-1.50mm}$^{18}$de Divitiis, N., ``Effect of Microlift Force on the Performance of Ultralight Aircraft'',
           {\it Journal of Aircraft}, Vol. 39, No. 2, 2002, pp. 318-325.

\medskip

\no  \hspace*{-1.50mm}$^{19}$ Lippisch A.M., ``Der 'Aerodynamische Bodeneffekt' und die Entwicklung des  Flugfl$\ddot{a}$chen-(Aerofoil-)Bootes'',
 { \it Luftfahrttechnik Raumfahrttechnik}, Vol. 10, 1964, pp.261-269.

\medskip

\no \hspace*{-1.50mm}$^{20}$Analytical Methods Inc. VSAERO User's Manual.
           Revision E5, April, 1994, pp. 1-52.

\newpage

\begin{center}

---------------------------------------------------
\vspace{-8.mm}

---------------------------------------------------
\vspace{-4.8mm}

{ Table 1. Dimensions and mass}

\vspace{-4.mm}
---------------------------------------------------
\vspace{-0.mm}

\end{center}

\begin{center}
 \begin{tabular}{lr}
Overall Length, $l$     ($m$)                           & 6  \\\\
Wing Span,      $b$     ($m$)                           & 6.8   \\\\
Planform Area,  $S$     ($m^2$)                         & 16   \\\\
Maximum Power,  $\Pi_{max}$ ($kw$)                          & 200  \\\\
Overall Weight, $W$     ($N$)                           & 9806 \\\\
\end{tabular}

\vspace{-4.mm}
---------------------------------------------------

\vspace{-8.mm}
---------------------------------------------------

\end{center}

\newpage

\begin{center}
{Table 2. Dimensionless aerodynamic derivatives vs. $h/c$. The values in the parenthesis are  from Ref. 13}
\end{center}

\begin{table}
  \begin{center}

  ---------------------------------------------------------------------------------------
\vspace{-4.mm}

  ---------------------------------------------------------------------------------------
\vspace{-0.mm}

  ---------------------------------------------------------------------------------------
\vspace{-6.mm}

  \begin{tabular}{lccccccccccc}
    $h/c$                 &  $ \hspace{1.cm}  \infty$ &     & {\hspace{1.cm} 1.0}        &        &  {\hspace{1.cm} 0.4}     &       \\[1pt]
    ${C_L}_\alpha$        & (3.0)     &  3.47  & (3.2)      &  3.7   & (3.6)   & 4.02  \\
    ${C_L}_h$             & (0.0)     & 0.0    & (-0.035)   & -0.042 & (-0.35) & -0.38 \\
    ${C_m}_\alpha$        & (-0.68)   & -0.7   & (-0.70)    &  -0.73 & (-0.73) & -0.76 \\
    ${C_m}_h$             & (0.0)     & 0.0    & (0.003)    & 0.0037 & (0.055) & 0.063 \\
    ${C_m}_q$             & (-4.44)   &-4.20   & (-4.44)    & -4.03  & (-4.44) & -3.97  \\
  \end{tabular}

 -----------------------------------------------------------------------------------------
\vspace{-4.mm}

 -----------------------------------------------------------------------------------------
\vspace{-0.mm}

  \end{center}
\end{table}

\newpage

\centerline { \bf List of captions }

\bigskip
\bigskip
\bigskip
\bigskip

\centerline {Fig. 1 Influence of $\varphi$ and $\vartheta$ on a vehicle in ground effect.}

\bigskip
\bigskip
\bigskip
\bigskip

\centerline {Fig. 2 Comparison of the results.}
\no a) Delta wing in dynamic ground effect. $AR = 1.456$, $\alpha = 22.1^\circ$, $\gamma = -10$. Symbols are from Ref. 7. \hspace{5.mm}
b), c), d) Delta wing in ground effect at different $\gamma$.
$AR = 1.456$, $\alpha = 10^\circ$. \hspace{5.mm}
e), f), g) Delta wings with different aspect ratios at $\alpha = 10^\circ$, $\gamma = -20^\circ$. Symbols are from Ref. 8.

\bigskip
\bigskip
\bigskip
\bigskip

\centerline {Fig. 3 The reference vehicle.}

\bigskip
\bigskip
\bigskip
\bigskip

\no Fig. 4 Vehicle lift coefficient and aerodynamic efficiency in ground effect obtained
for $h/c$ = $0.167$, $0.614$, $2.792$, $\infty$, at $\varphi$ = $\beta$ = 0.

\bigskip
\bigskip
\bigskip
\bigskip

\centerline {Fig. 5 Influence of pitch angle on the aerodynamic coefficients.
$h/c$ = 1, $\gamma$ = $\beta$ = $\varphi$ = 0.}

\bigskip
\bigskip
\bigskip
\bigskip

\centerline {Fig. 6 Influence of roll angle on the aerodynamic coefficients.
$\alpha$ = 5$^\circ$, $\beta$ = $\gamma$ = 0.}

\bigskip
\bigskip
\bigskip
\bigskip

\centerline {Fig. 7 Influence of flight path angle on the aerodynamic coefficients.
$\alpha$ = 5$^\circ$, $\beta$ = $\varphi$ = 0.}

\bigskip
\bigskip
\bigskip
\bigskip

\no Fig. 8 Vehicle rotational derivatives in ground effect in function of the angle of attack. $\beta$ = $\varphi$ = 0.

\bigskip
\bigskip
\bigskip
\bigskip

\no Fig. 9 Comparison of the dimensionless aerodynamic derivatives for $h/c = 0.1$.
Data from Ref. 5 in horizontal axis, present results in vertical axis.

\bigskip
\bigskip
\bigskip
\bigskip

\centerline {Fig. 10 Trim calculation in horizontal flight at different heights. $\beta$ = $\varphi$ = 0.}

\bigskip
\bigskip
\bigskip
\bigskip

\centerline {Fig. 11 Influence of the attitude on the trim in horizontal flight.}

\bigskip
\bigskip
\bigskip
\bigskip

\centerline {Fig. 12  Trim analysis of a "truly banked" turn with $\zeta$ = 5$^\circ$ s$^{-1}$ }

\bigskip
\bigskip
\bigskip
\bigskip

\centerline {Fig. 13 Root Locus.   ${\bf H}=0$ a) and b),   ${\bf H} \ne 0$ c) and d).}

\bigskip
\bigskip
\bigskip
\bigskip

\no {Fig. 14 Time history of the perturbed longitudinal motion beginning from different initial heights, (${\bf H} \ne 0$).}

\bigskip
\bigskip
\bigskip
\bigskip

\no  {Fig. 15 Time history of the perturbed lateral motion beginning from different initial heights, (${\bf H} \ne 0$).}

\newpage

\begin{figure}[!ht]
\begin{center}
\psfig{file=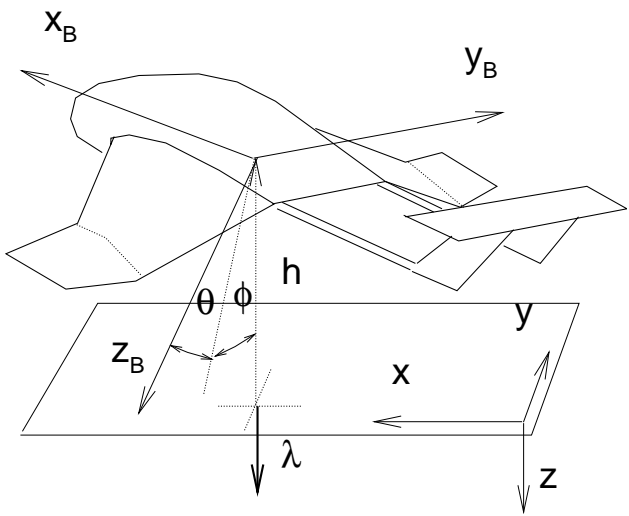,rheight=0.0cm}
\vspace{0.mm}
\caption{Influence of $\varphi$ and $\vartheta$ on a vehicle in ground effect.}
\end{center}
\end{figure}
\newpage

\begin{figure}[!ht]
\begin{center}
\psfig{file=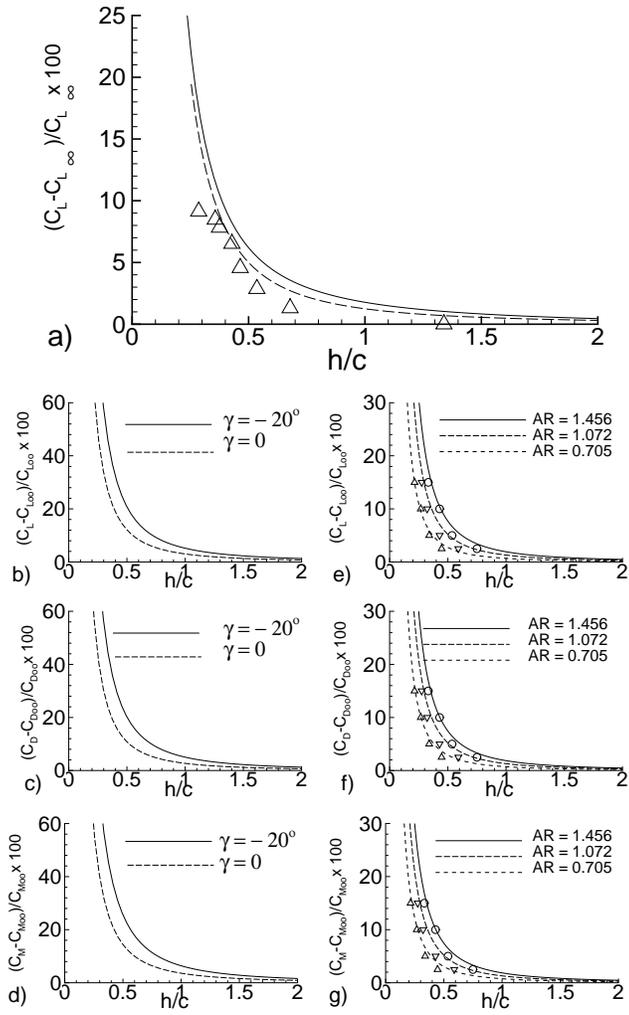,rheight=0.0cm}
\vspace{0.mm}
\caption{Comparison of the results. a) Delta wing in dynamic ground effect. $AR = 1.456$, 
$\alpha = 22.1^\circ$, $\gamma = -10$. Symbols are from Ref. 7. \hspace{5.mm}
b), c), d) Delta wing in ground effect at different $\gamma$.
$AR = 1.456$, $\alpha = 10^\circ$. \hspace{5.mm}
e), f), g) Delta wings with different aspect ratios at $\alpha = 10^\circ$, $\gamma = -20^\circ$. Symbols are from Ref. 8.}
\end{center}
\end{figure}
\newpage

\begin{figure}[!ht]
\begin{center}
\psfig{file=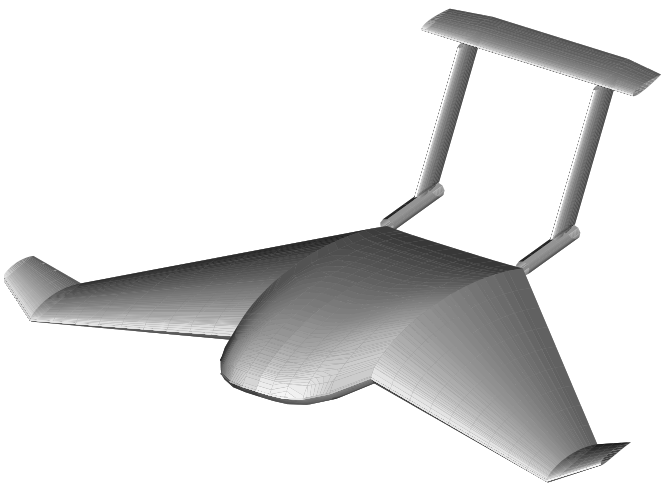,rheight=0.0cm}
\vspace{0.mm}
\caption{The reference vehicle.}
\end{center}
\end{figure}
\newpage

\begin{figure}[!ht]
\begin{center}
\psfig{file=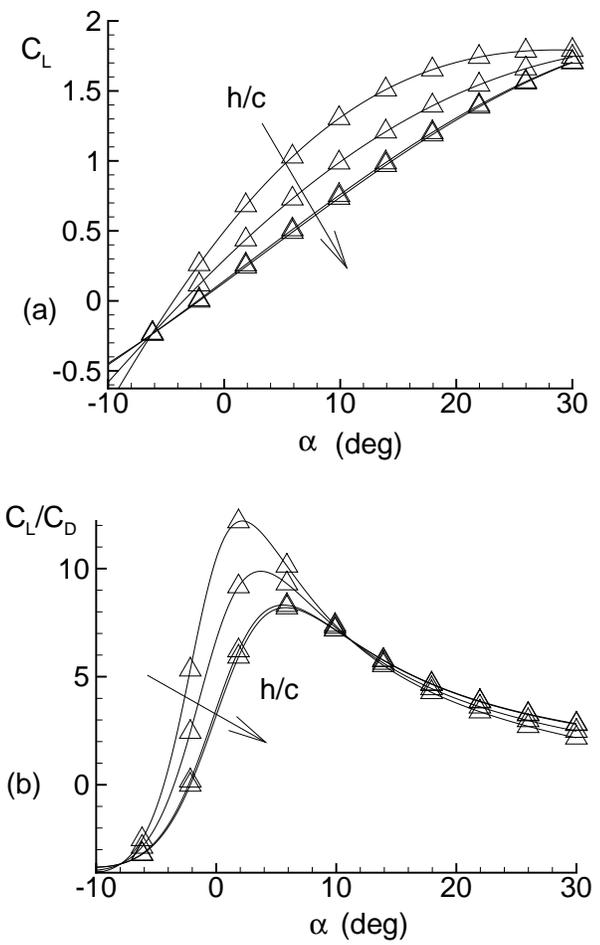,rheight=0.0cm}
\vspace{0.mm}
\caption{Vehicle lift coefficient and aerodynamic efficiency in ground effect obtained
for $h/c$ = $0.167$, $0.614$, $2.792$, $\infty$, at $\varphi$ = $\beta$ = 0.}
\end{center}
\end{figure}
\newpage

\begin{figure}[!ht]
\begin{center}
\psfig{file=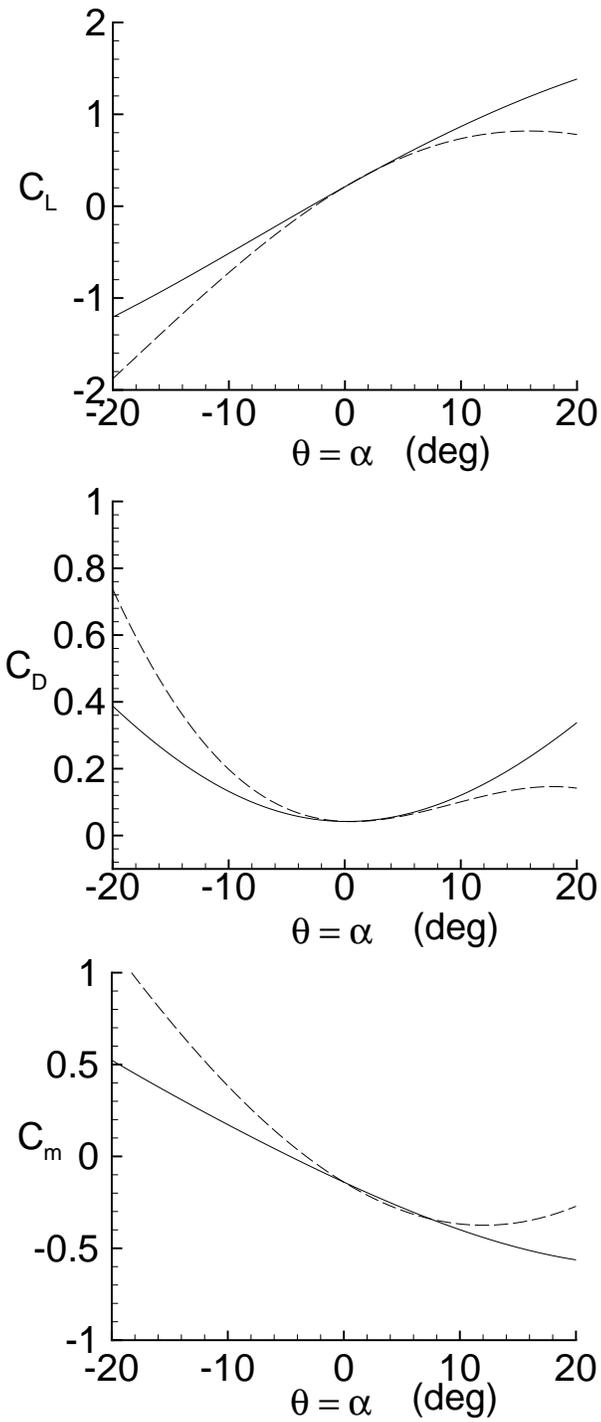,rheight=0.0cm}
\vspace{0.mm}
\caption{Influence of pitch angle on the aerodynamic coefficients.
$h/c$ = 1, $\gamma$ = $\beta$ = $\varphi$ = 0.}
\end{center}
\end{figure}
\newpage

\begin{figure}[!ht]
\begin{center}
\psfig{file=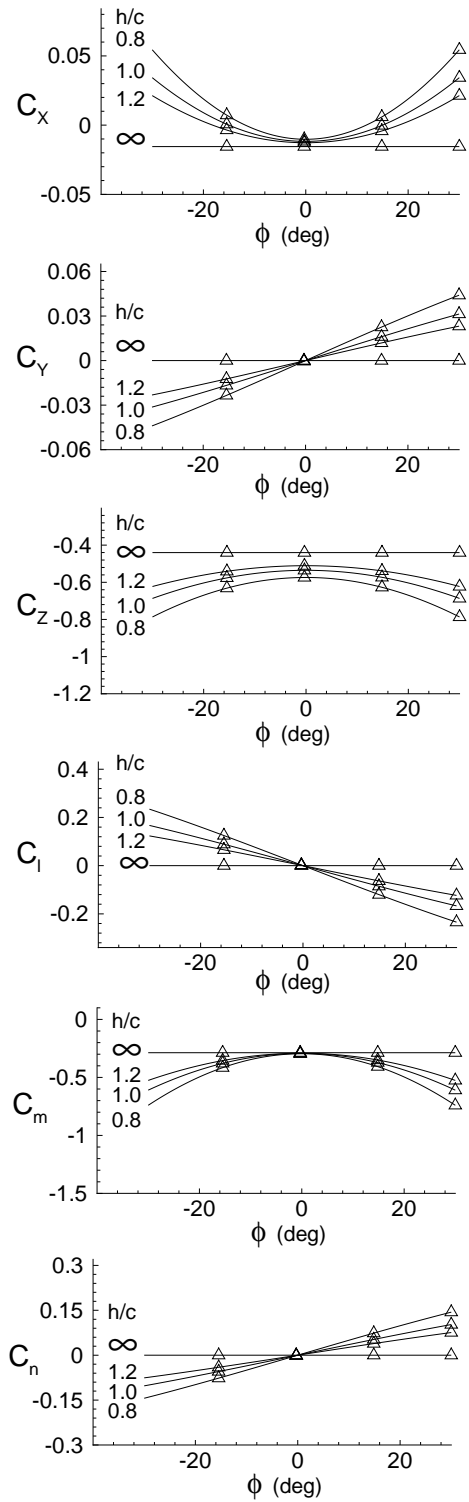,rheight=0.0cm}
\vspace{0.mm}
\caption{Influence of roll angle on the aerodynamic coefficients.
$\alpha$ = 5$^\circ$, $\beta$ = $\gamma$ = 0.}
\end{center}
\end{figure}
\newpage

\begin{figure}[!ht]
\begin{center}
\psfig{file=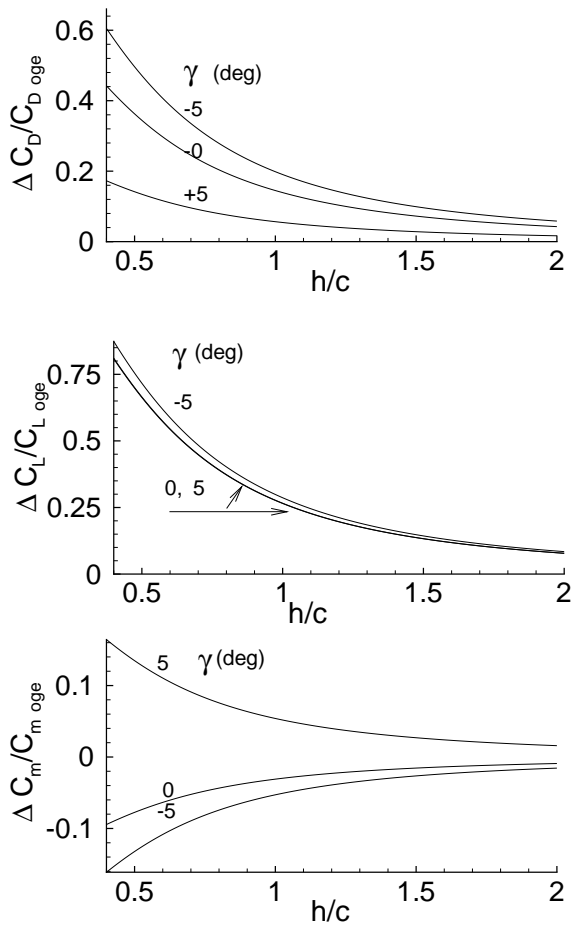,rheight=0.0cm}
\vspace{0.mm}
\caption{Influence of flight path angle on the aerodynamic coefficients.
$\alpha$ = 5$^\circ$, $\beta$ = $\varphi$ = 0.}
\end{center}
\end{figure}
\newpage

\begin{figure}[!ht]
\begin{center}
\psfig{file=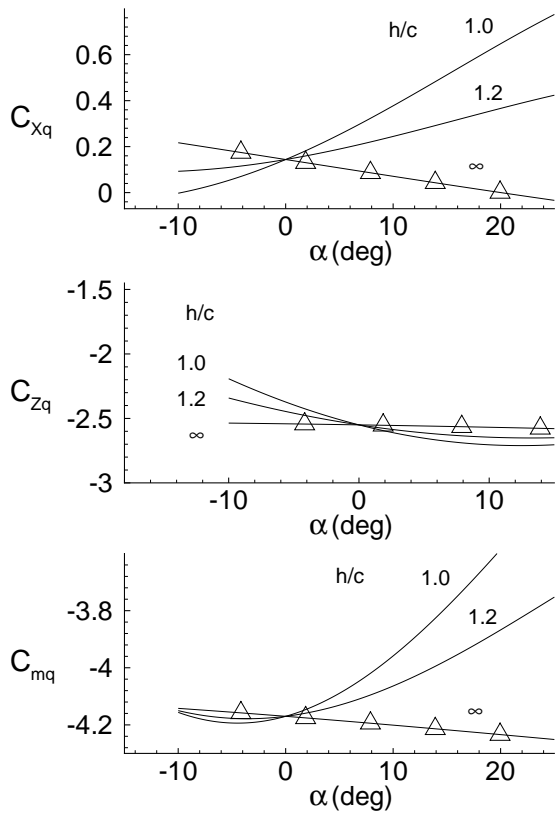,rheight=0.0cm}
\vspace{0.mm}
\caption{Vehicle rotational derivatives in ground effect in function of the angle of attack. $\beta$ = $\varphi$ = 0.}
\end{center}
\end{figure}
\newpage

\begin{figure}[!ht]
\begin{center}
\psfig{file=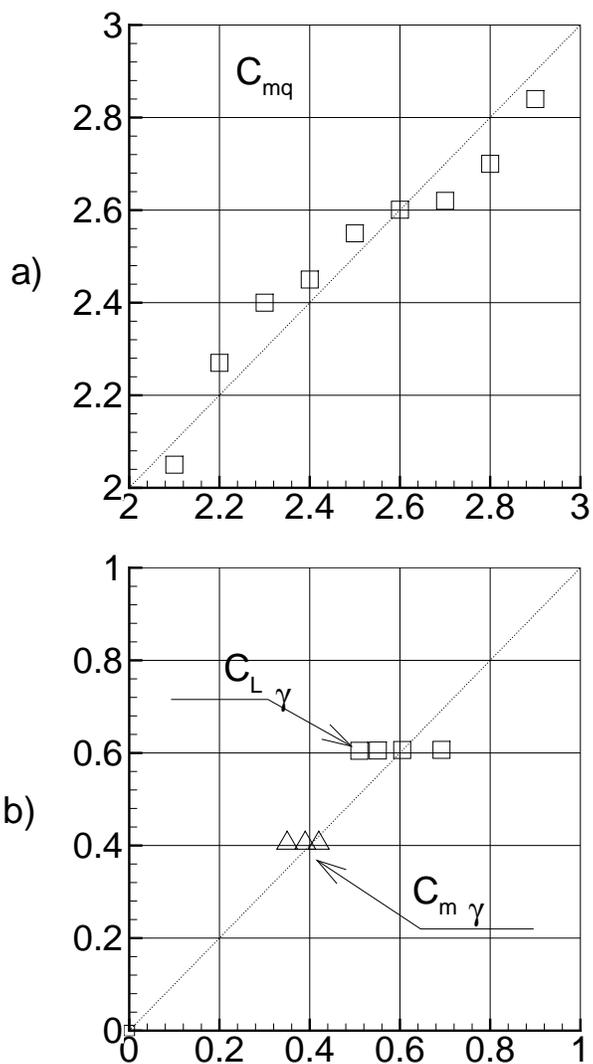,rheight=0.0cm}
\vspace{0.mm}
\caption{Comparison of the dimensionless aerodynamic derivatives for $h/c = 0.1$.
Data from Ref. 5 in horizontal axis, present results in vertical axis.}
\end{center}
\end{figure}
\newpage

\begin{figure}[!ht]
\begin{center}
\psfig{file=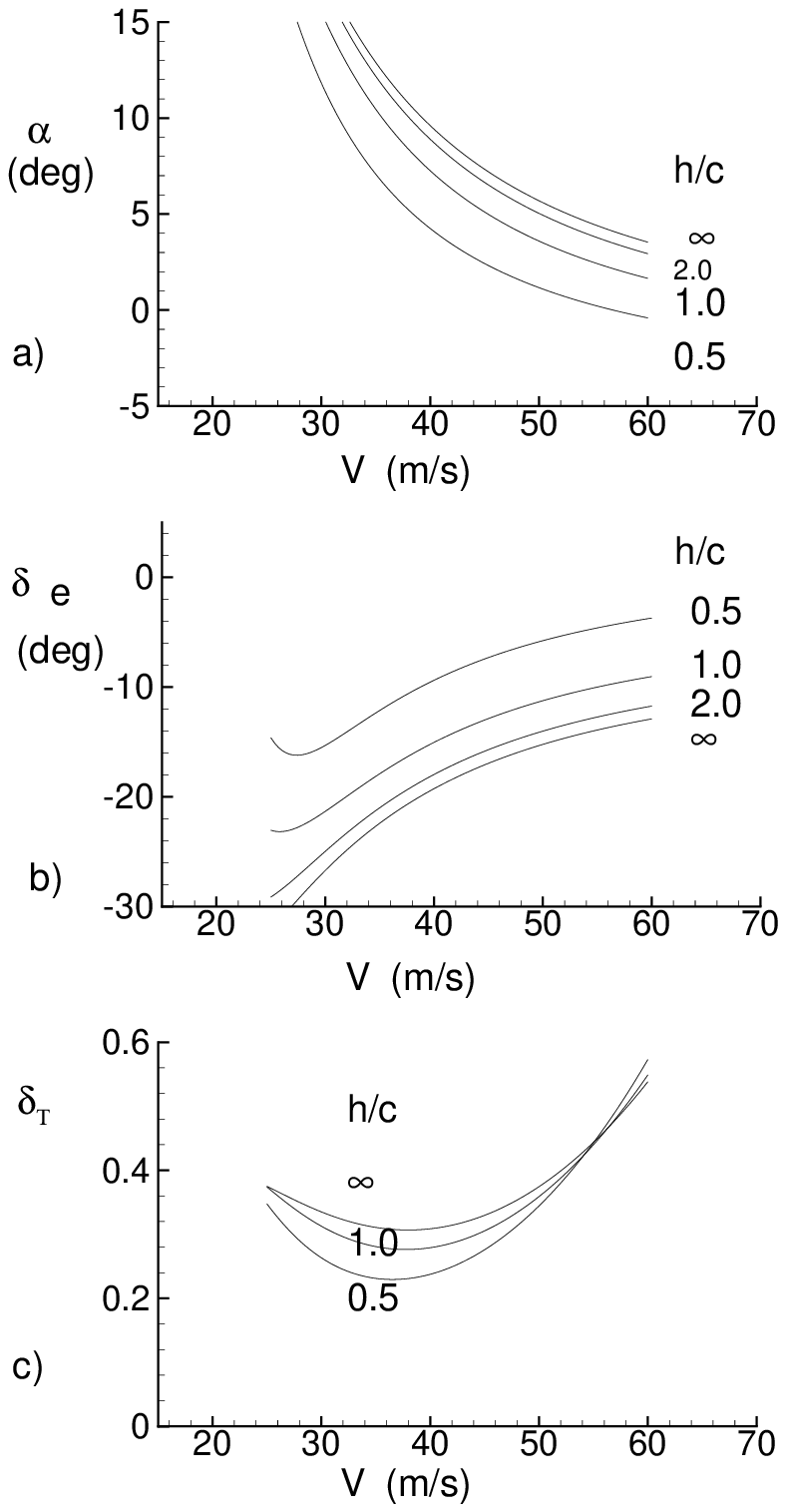,rheight=0.0cm}
\vspace{0.mm}
\caption{Trim calculation in horizontal flight at different heights. $\beta$ = $\varphi$ = 0.}
\end{center}
\end{figure}
\newpage

\begin{figure}[!ht]
\begin{center}
\psfig{file=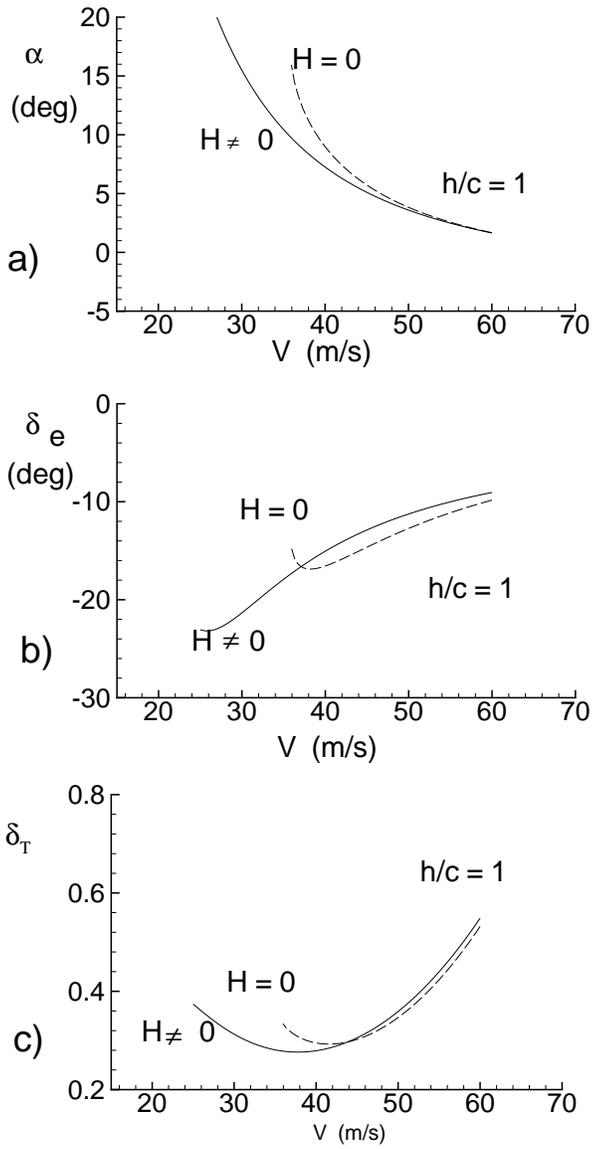,rheight=0.0cm}
\vspace{0.mm}
\caption{Influence of the attitude on the trim in horizontal flight.}
\end{center}
\end{figure}
\newpage

\begin{figure}[!ht]
\begin{center}
\psfig{file=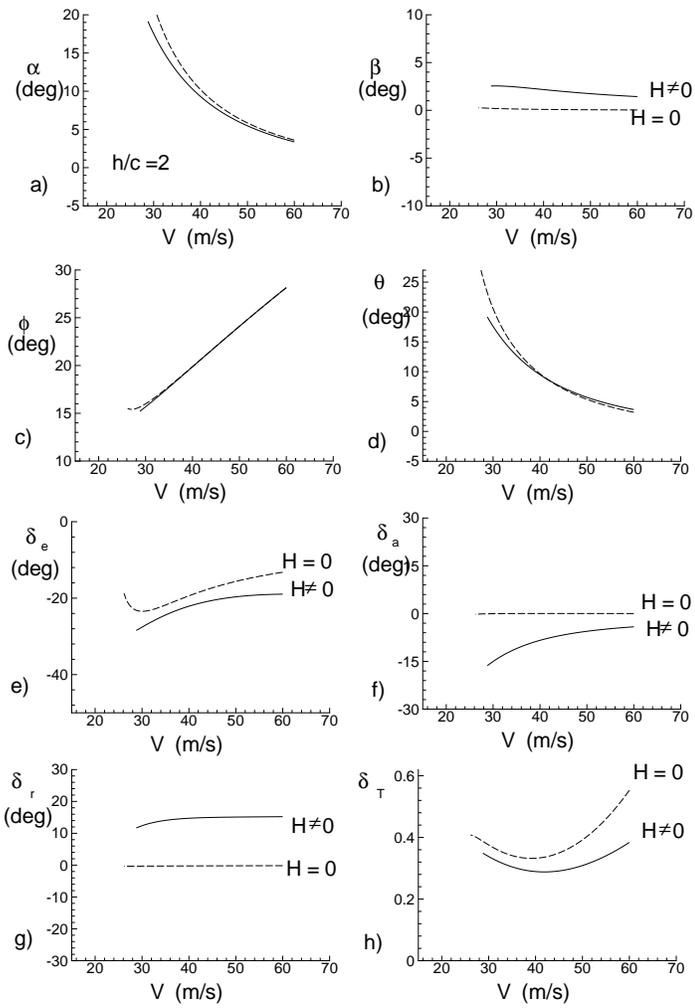,rheight=0.0cm}
\vspace{0.mm}
\caption{Trim analysis of a "truly banked" turn with $\zeta$ = 5$^\circ$ s$^{-1}$ }
\end{center}
\end{figure}
\newpage

\begin{figure}[!ht]
\begin{center}
\psfig{file=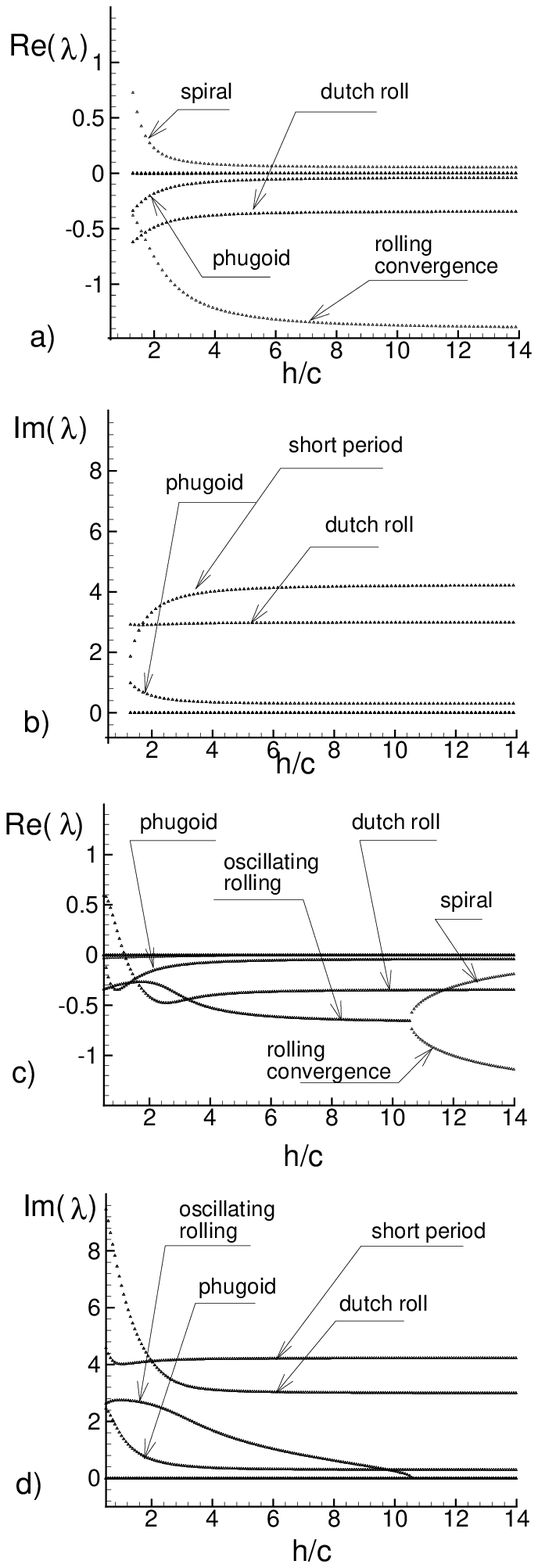,rheight=0.0cm}
\vspace{0.mm}
\caption{Root Locus.   ${\bf H}=0$ a) and b),   ${\bf H} \ne 0$ c) and d).}
\end{center}
\end{figure}
\newpage

\begin{figure}[!ht]
\begin{center}
\psfig{file=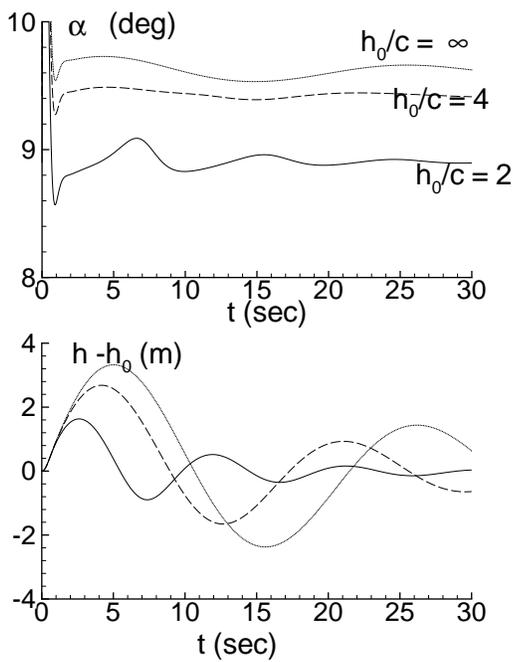,rheight=0.0cm}
\vspace{0.mm}
\caption{ Time history of the perturbed longitudinal motion beginning from different initial heights, (${\bf H} \ne 0$).}
\end{center}
\end{figure}
\newpage

\begin{figure}[!ht]
\begin{center}
\psfig{file=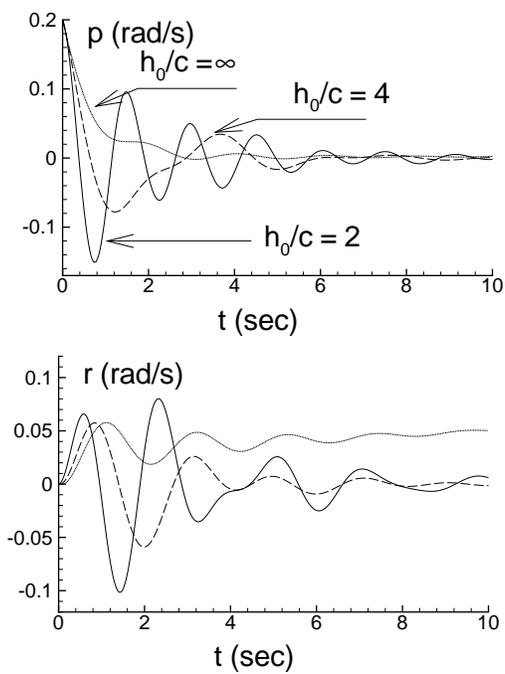,rheight=0.0cm}
\vspace{0.mm}
\caption{Time history of the perturbed lateral motion beginning from different initial heights, (${\bf H} \ne 0$).}
\end{center}
\end{figure}
\newpage

\end{document}